\documentclass[prc,unsortedaddress,superscriptaddress,amsmath,amsfonts,amstex,amssymb,showpacs,floatfix]{revtex4}
\usepackage{graphics,epsfig}
\begin{document}

\title{ Astrophysical $S$ factor  for the ${}^{15}{\rm N}(p,\gamma){}^{16}{\rm O}$ reaction  from $R$-matrix analysis
and asymptotic normalization coefficient for ${}^{16}{\rm O} \to {}^{15}{\rm N} + p$. Is any fit acceptable?}

\author{A.\,M.~Mukhamedzhanov}
\affiliation{Cyclotron Institute, Texas A\&M University,
College Station, TX 77843}

\author{M.~La Cognata}
\affiliation{Universit\`a di Catania and INFN Laboratori Nazionali
del Sud, Catania, Italy}

\author{V.~Kroha}
\affiliation{Nuclear Physics Institute, Czech Academy of Sciences,
250 68 \v{R}e\v{z} near Prague,
Czech Republic}

\begin{abstract}
The $^{15}{\rm N}(p,\gamma)^{16}{\rm O}$ reaction provides a path from the CN cycle to the CNO bi-cycle and CNO tri-cycle. The measured  astrophysical factor for this reaction is dominated by resonant capture through two strong $J^{\pi}=1^{-}$ resonances at $E_{R}= 312$ and 
$962$ keV and direct capture to the ground state. Recently, a new measurement of the astrophysical factor 
for the $^{15}{\rm N}(p,\gamma)^{16}{\rm O}$ reaction has been published [P. J. LeBlanc {\it et al.}, Phys. Rev. {\bf C 82}, 055804 (2010)].  The analysis has been done using the $R$-matrix approach with unconstrained variation of all parameters including the asymptotic normalization coefficient (ANC). The best fit has been obtained for the square of the ANC  $C^{2}= 539.2$ fm${}^{-1}$, which exceeds the previously measured value by a factor of $\approx 3$. Here we present a new $R$-matrix analysis of the Notre Dame-LUNA data with the fixed within the experimental uncertainties square of the ANC $C^{2}=200.34$ fm${}^{-1}$. Rather than varying the ANC we add the contribution from a background resonance that effectively takes into account contributions from higher levels. Altogether we present 8 fits, five unconstrained and three constrained. In all the fits the ANC is fixed at the previously determined experimental value $C^{2}=200.34$ fm${}^{-1}$.
For the unconstrained fit with the boundary condition $B_{c}=S_{c}(E_{2})$, where $E_{2}$ is the energy of the second level, we get $S(0)=39.0 \pm 1.1 $ keVb and normalized ${\tilde \chi}^{2}=1.84$, i.e. the result which is similar to [P. J. LeBlanc {\it et al.}, Phys. Rev. {\bf C 82}, 055804 (2010)]. From all our fits we get the range $33.1 \leq S(0) \leq 40.1$ keVb which overlaps with the result of [P. J. LeBlanc {\it et al.}, Phys. Rev. {\bf C 82}, 055804 (2010)]. We address also physical interpretation of the fitting parameters.   
\end{abstract}

\pacs{26.20.-f, 25.60.Tv, 24.30.-v, 29.85.-c}

\maketitle

\section{Introduction}
The $^{15}{\rm N}(p,\gamma)^{16}{\rm O}$ reaction provides the path to form $^{16}{\rm O}$ in stellar hydrogen burning, thus transforming the CN cycle into the CNO bi-cycle and CNO tri-cycle. In stellar environments, the $^{15}{\rm N}(p,\gamma)^{16}{\rm O}$ reaction proceeds at very low energies, where it is dominated by resonant capture to the ground state through the first two interfering $J^{\pi}=1^{-}$ $\,s$ wave resonances at $E_{R}=312$ and $964$ keV, where $E_{R}$ is the resonance energy in the center of mass (c. m.) system. There is also a small contribution from direct capture to the ground state of ${}^{16}{\rm O}$, which turns out to play an important role due to the interference with the resonant amplitudes.

In our previous paper \cite{muk08}, the measurement of the asymptotic normalization coefficient (ANC) 
for the ${}^{16}{\rm O} \to {}^{15}{\rm N} + p$ has been reported. This ANC has been used to fix the non-resonant contribution to the $^{15}{\rm N}(p,\gamma)^{16}{\rm O}$ capture and we found that it was impossible to fit the low-energy data from \cite{rolfs74}. Moreover, we underscored that to fit these experimental data one needs to increase the ANC almost by on order of magnitude. Our calculated astrophysical factor using the two-level, two-channel $R$-matrix approach led to $S(0)= 36.0 \pm  
6.0$ keVb, which is significantly smaller than $S(0)=64 \pm  6$ keVb reported in \cite{rolfs74} but in agreement  with the older measurements in \cite{hebb60}. Correspondingly,  
we have found that for every $2200 \pm 300$ cycles of the main CN cycle, one CN catalyst is lost due to the $^{15}{\rm N}(p,\gamma)^{16}{\rm O}$ reaction, rather than $1200 \pm 100$ cycles determined from data of Ref. \cite{rolfs74}. Our results were confirmed later by \cite{barker08}. Recently, two new measurements of the astrophysical factor for the  $^{15}{\rm N}(p,\gamma)^{16}{\rm O}$ have been published. The first measurement was performed at the LUNA underground accelerator facility at the Gran Sasso laboratory \cite{bemmerer}. This measurement covered only the low-energy region, $E \leq 230$ keV, where $E$ is the relative ${}^{15}{\rm N}-p$ energy. The second study of $^{15}{\rm N}(p,\gamma)^{16}{\rm O}$, which has been just reported in \cite{LeBlanc}, was performed over a wide energy range at the Notre Dame Nuclear Science Laboratory (NSL) and the LUNA II facility. The obtained $S(0)= 39.6 \pm 2.6$ keVb is in a perfect agreement with our prediction $S(0)=36.0 \pm 6.0$ keVb \cite{muk08}. However, in the $R$-matrix fitting  of the experimental data the ANC was used as an unconstrained fitting parameter and the best fit with the normalized ${\tilde \chi}^{2}= 1.80$ has been achieved for the square of the ANC $C^{2}= 539 \pm 138$ fm${}^{-1}$, which is significantly higher then our measured value $C^{2}= 192 \pm 26$ fm${}^{-1}$. 
The ANC is a fundamental nuclear characteristics \cite{blokhintsev} and, if it is available from independent experimental measurements, it can be varied only within experimental boundaries. An unconstrained variation of the ANC to achieve the best fit might signal that some physical input is missing in the reaction model and this incomplete knowledge is compensated for by adopting an unphysical value of the ANC. In the case under consideration, in the $R$-matrix approach the ANC determines the overall normalization of the non-resonant radiative capture amplitude and the channel (external) part of the radiative width amplitudes of the both resonances involved. Although these amplitudes are small and any sizable impact on the astrophysical factor can be achieved only by a significant variation of the ANC, the contribution of the non-resonant amplitude increases toward low energies, which is the region of the astrophysical interest. However, not every best fit can be accepted if physical parameters of the fit exceed the previously well established experimental limits.
A similar situation has occurred in the analysis of the ${}^{14}{\rm N}(p,\gamma){}^{15}{\rm O}$ reaction for transition to the ground state, where the non-resonant capture contribution is also controlled by the ANC 
for the ${}^{15}{\rm O}({\rm gr. st.}) \to {}^{14}{\rm N} + p$ \cite{muk2003,solarfusion}. The best fit
is achieved at $C \approx 11$ fm${}^{-1/2}$, but this fit was not accepted because the recommended ANC value is $7.4 \pm 0.5$ fm${}^{-1/2}$ \cite{muk2003, solarfusion}.  The recommended in \cite{solarfusion} $S(0)=0.27 \pm 0.05$ keVb factor corresponds to the fit with the recommended ANC but the expanded uncertainty is not due to the best fit at higher ANC but to a possible contribution of the capture to the channel spin $I=1/2$, which interferes with  
the 259 keV resonance. It is an instructive example when missing physics is taken into account rather than an
unconstrained variation of the ANC, which demonstrates that it is important to keep track on the boundaries of  variation of the fitting parameters in accordance with the previously available information about them obtained from other sources. For example, in the case under consideration these additional parameters are the observable partial resonance widths in the proton and $\alpha$-channels, which have been determined from the $R$-matrix analysis \cite{barker08,lacognata09} of the ${}^{15}{\rm N}(p,\alpha){}^{16}{\rm C}$  data \cite{rolfs82}. It is especially important because eventually the results of the fit are aimed to deliver vital astrophysical information and to make it reliable all available nuclear physics information should be invoked.

In this paper, we present our own fits of the data from \cite{LeBlanc} with the fixed experimental ANC 
but adding the contribution from a background resonance. Results of our fit are compared with the one presented in \cite{LeBlanc} obtained using the AZURE $R$-matrix code \cite{azuma}.  Besides, we compare the parameters of both fits and present also the fit where observable particle widths are consistent with the previous fits of the ${}^{15}{\rm N}(p,\alpha){}^{16}{\rm C}$  reaction.

\section{ANC}
We start with the physical meaning of the ANC.  The residue of the scattering $S$ matrix in the corresponding bound-state pole is expressed in terms of the ANC \cite{muk77}, what provides a model-independent definition of the ANC. In the $R$-matrix approach the ANC determines the normalization of the external non-resonant radiative capture amplitude and the channel radiative reduced width amplitude \cite{tang}. In a single-particle approach, the nucleon ANC can be expressed in terms of the spectroscopic factor and the 
single-particle bound-state wave function of the nucleon calculated in some adopted mean-field:
\begin{equation} 
C^{2}= S\,b^{2},
\label{singlpart1}
\end{equation}
where $S$ is the spectroscopic factor and $b$ is the single-particle ANC, i.e. the amplitude of the single-particle bound-state wave function. Note that the isospin Clebsch-Gordan coefficient is absorbed in the 
spectroscopic factor.
In such an approach we can consider the ground state of ${}^{16}{\rm O}$ as the bound state $({}^{15}{\rm N}\,p)$ with the proton occupying the single-particle orbital $1p_{1/2}$. The spectroscopic factor shows the probability of this configuration in the ground state of ${}^{16}{\rm O}$. Due to the identity of nucleons this probability can be larger than one. In a simple independent particle shell model the spectroscopic factor of the $1p_{1/2}$ state is equal to the number of protons occupying this orbital, i.e. 2. To determine the spectroscopic factor from Eq. (\ref{singlpart1})
one needs to determine the proton bound-state wave function. To this end  we adopt the Woods-Saxon potential with the standard geometry, $r_{0}= 1.25$ fm and diffuseness $a=0.65$ fm and the spin-orbit potential depth of $6.39$ MeV. Assuming that the proton is in the $p_{1/2}$ orbital we obtain the single-particle proton ANC $b \approx 9.96$ fm${}^{-1/2}$. If one adopts the ANC from \cite{LeBlanc}, we obtain the spectroscopic factor $S=5.44$ versus $S=2.1$ obtained from our $C^{2}= 192$ fm${}^{-1}$. Even if we adopt an unrealistically large radius $r_{0}=1.50$ fm and $a=0.65$ fm for the Woods-Saxon potential, we obtain $b=13.6$ fm${}^{-1/2}$ and too high spectroscopic factor $S=2.81$. 

Definitely such a high spectroscopic factor obtained from the ANC adopted in \cite{LeBlanc} requires a physical interpretation. 
In this aspect it would be useful to present the phenomenological spectroscopic factors obtained from the analysis of different reactions. For example, the spectroscopic factor deduced from the analysis of ${}^{16}{\rm O}(e,e'\,p){}^{15}{\rm N}$ reaction is $S=1.27 \pm 0.13$ \cite{Leuschner,kramer}. The proton bound-state wave function deduced from the $(e,e'\,p)$ reaction is reproduced by the Woods-Saxon potential with the geometry
$r_{0}=1.37$ fm and $a=0.65$ fm. The single-particle ANC of this bound-state wave function is $b=11.62$ 
fm${}^{-1/2}$. Using the upper limit of the deduced spectroscopic factor $S=1.40$ we obtain the square of the 
proton ANC in the ground state of ${}^{16}{\rm O}$ $\,C^{2}= 189$ fm${}^{-1}$, which is in a perfect agreement with our result \cite{muk08}. In our paper \cite{muk08}, we presented the spectroscopic factors extracted from different ${}^{15}{\rm N}({}^{3}{\rm He},d){}^{16}{\rm O}$ reactions
including our result. Besides our spectroscopic factor $S=2.1$ \cite{muk08},  three other measurements (see references in \cite{muk08}) gave $S \leq 1.76$. The DWBA reanalysis of the ${}^{16}{\rm O}({}^{3}{d,\,\rm He}){}^{15}{\rm N}$ reaction \cite{Hiebert} performed in \cite{kramer} using the proton bound-state wave function obtained from the ${}^{16}{\rm O}(e,e'\,p){}^{15}{\rm N}$ reaction with $r_{0}= 1.37$ fm and $a=0.65$ fm gave even lower spectroscopic factor $S=1.02$. With this low spectroscopic factor we get $C^{2} \approx 140$ fm$^{-1}$.  
Since the measurements in \cite{Hiebert} were done in 1967, the accuracy of the absolute normalization of the differential cross section in \cite{Hiebert} might be questionable. Spectroscopic factors below 2 have also been obtained in microscopic calculations \cite{Geurts,Jensen}. Concluding this discussion we cannot find  any justification for such a high value of the ANC used in \cite{LeBlanc} because it leads to an unphysical spectroscopic factor. 
We only can assume that a broad variation of the ANC beyond of the experimental limits has been used in \cite{LeBlanc} to compensate for missing mechanisms in the reaction model.     

\section{$R$ matrix}
To underscore the role of the ANC in the fitting experimental data in this section we present the expression for the astrophysical factor in the $R$-matrix approach, which we use for the analysis of the experimental   data  \cite{LeBlanc}. 
It is two-level, two-channel  $R$-matrix, which includes the coherent contribution from two $1^{-}$ resonances and  non-resonant term describing  direct capture. The ANC determines the normalization of the non-resonant capture amplitude, which describes the external direct capture in the $R$-matrix approach and the channel radiative width amplitude, which are important for the fitting.
But, in addition to \cite{LeBlanc}, we add the coherent contribution from a background resonance rather than varying  the ANC. We can assume that this background pole takes effectively into account  contributions from distant levels with $J^{\pi}=1^{-}$. 
The expression for the astrophysical factor in the $R$-matrix method for the case under consideration can be written as \cite{lanethomas58,barkerkajino}
\begin{align}
S(E)(keVb) = \frac{\pi\,\lambda_{N}^{2}}{2}\,\frac{ {\hat J}_{R} }{ {\hat J}_{x}\,{\hat J}_{A} }\,\frac{m_{x}+m_{A} }{m_{x}\,m_{A}}\,{931.5}^{2}\,e^{2\,\pi\,\eta}\,10\,[ \sum\limits_{\nu ,\tau  = 1,2} ({\Gamma _{\nu\, \gamma {\mathop{\rm (int)}} }^{1/2} \pm \Gamma _{\nu \,\gamma {\mathop{\rm ext}}}^{1/2})\,{[{\bold A}^{-1}]_{\nu \tau }}} \Gamma _{\tau\, p}^{1/2}\,\, \pm \,\,{M_{DC}} + \,{M_{BG}}],
\label{Sfactor1}
\end{align}
where $\lambda_{N}= 0.2118$ fm is the nucleon Compton wave length, $931.5$ is the atomic mass unit in MeV, $Z_{j}$ and $m_{j}$ are the charge and mass of particle $j$ and $\mu_{ij}$ is the reduced mass of particles $i$ and $j$,
$\eta$ is the Coulomb parameter in the initial state of the reaction, $J_{j}$ is the spin of particle $j$ and $J_{R}$ is the spin of the resonance, ${\hat J}= 2\,J + 1$, $\,\,k_{\gamma}= (E+ \varepsilon)/(\hbar\,c)$ is the momentum of the emitted photon expressed in fm${}^{-1}$, $E$ is the relative $p-A$ energy, $A$ is the target, $\varepsilon$ is the proton binding energy of the bound state $(A\,p)$, $L$ is the multipolarity of the electromagnetic transition, ${\bold A}$ is the standard level matrix for the two-channel, two-level case \cite{lanethomas58}, 
$\Gamma_{\tau c}^{1/2}= \sqrt{2\,P_{l_{c}}(k_{c},r_{c})}\,\gamma_{\tau\, c}$, $\,\gamma_{\tau \,c}$ is the reduced width amplitude for the level $\tau=1,2$ in the channel $c=p,\alpha$,  $\,P_{l_{c}}(k_{c},r_{c})$ is the barrier
penetrability factor in the channel $c$, $l_{c}$ is the orbital angular momentum of the resonance and  $k_{c}$ is the relative momentum of the particles in the channel $c$, $r_{c}$ is the $R$-matrix channel radius in the channel $c$, $\,\Gamma _{\nu \, \gamma {\mathop{\rm (int)}} }^{1/2}= \sqrt{2}\,k_{\gamma}^{L+ 1/2}\,\gamma_{\nu \,\gamma {\mathop{\rm (int)}} }$, $\,\gamma_{\nu \, \gamma {\mathop{\rm (int)}}}$ is the internal radiative reduced width amplitude for transition from the level $\nu$ to a bound state (the ground state in the case under consideration),
$\Gamma _{\nu \, \gamma ext}^{1/2}= \sqrt{2}\,k_{\gamma}^{L+ 1/2}\,\gamma_{\nu \, \gamma {\mathop{\rm ext}}}$, $\,\gamma_{\nu \, \gamma {\mathop{\rm ext}}}$ is the complex channel (external) radiative reduced width amplitude for transition from the level $\nu$ to a bound state
given by the expression 
\begin{align}
\gamma_{\nu \, \gamma {\mathop{\rm (ext)}}}=  C\,\sqrt {\frac{1}{2}\,\frac{{{e^2}}}{{\hbar c}}{\lambda _N}\frac{{931.5}}{E}} \,\,r_p^{L + 1/2}\mu _{pA}^L[\frac{{{Z_p}}}{{m_p^L}} + {( - 1)^L}\frac{{{Z_A}}}{{m_A^L}}]\sqrt {\frac{{(L + 1){\hat L}}}{L}} \,\frac{1}{{{\hat L}!!}}\Gamma _{\nu \, p}^{1/2}\sqrt {{P_{{l_p}}}({k_{p,}}{r_p})}  \nonumber\\ 
 [{F_{l_{p}}^2}({k_p},{r_p})\,\,\, + \,{G_{l_{p}}^2}({k_p},{r_p})]{W_{ - {\eta _f},{l_f} + 1/2}}({r_p}) < {l_p}0\,\,L\,\,0|\,{l_f}0 >\, U(L{l_f}{J_i}I;\,\,{l_p}{J_f}) \nonumber\\
\times \lbrack In{t_1}\,\, + \,\,i\,\frac{{F_{l_{p}}({k_p},{r_p})G_{l_{p}}({k_p},{r_p})\,}}{{{k_p}{r_p}}}{P_{{l_p}}}({k_{p,}}{r_p})\,In{t_2}\rbrack, 
\label{gext1} 
\end{align}
\begin{align}
 In{t_1} = \frac{{{F^2}({k_p},{r_p})\,}}{{{k_p}{r_p}}}{P_{{l_p}}}({k_{p,}}{r_p})\int\limits_{{r_p}}^\infty  {dr} \frac{{{r^L}}}{{r_p^{L + 1}}}\frac{{{W_{ - {\eta _f},{l_f} + 1/2}}(r)}}{{{W_{ - {\eta _f},{l_f} + 1/2}}({r_p})}}\frac{{F_{l_{p}}({k_p},r)}}{{F_{l_{p}}({k_p},{r_p})}}   \nonumber\\ 
 + \frac{{{G_{l_{p}}^2}({k_p},{r_p})\,}}{{{k_p}{r_p}}}{P_{{l_p}}}({k_{p,}}{r_p})\int\limits_{{r_p}}^\infty  {dr} \frac{{{r^L}}}{{r_p^{L + 1}}}\frac{{{W_{ - {\eta _f},{l_f} + 1/2}}(r)}}{{{W_{ - {\eta _f},{l_f} + 1/2}}({r_p})}}\frac{{G_{l_{p}}({k_p},r)}}{{G_{l_{p}}({k_p},{r_p})}},  
\label{int1} 
\end{align}
\begin{align}
 In{t_2} = \int\limits_{{r_p}}^\infty  {dr} \frac{{{r^L}}}{{r_p^{L + 1}}}\frac{{{W_{ - {\eta _f},{l_f} + 1/2}}(r)}}{{{W_{ - {\eta _f},{l_f} + 1/2}}({r_p})}}\frac{{F_{l_{p}}({k_p},r)}}{{F_{l_{p}}({k_p},{r_p})}} 
 -\int\limits_{{r_p}}^\infty  {dr} \frac{{{r^L}}}{{r_p^{L + 1}}}\frac{{{W_{ - {\eta _f},{l_f} + 1/2}}(r)}}{{{W_{ - {\eta _f},{l_f} + 1/2}}({r_p})}}\frac{{G_{l_{p}}({k_p},r)}}{{G_{l_{p}}({k_p},{r_p})}},  
\label{int2} 
\end{align}
$F_{l_{p}}({k_p},{r})$ and  $G_{l_{p}}({k_p},{r_p})$ are the Coulomb regular and singular solutions,
$W_{ - \eta _{f},\,l_{f} + 1/2}(r)$ is the Whittaker function describing the radial dependence of the tail of the bound state wave function to which transition occurs after the photon is emitted, $\eta_{f}$ is the Coulomb parameter of the bound state and $l_{f}$ its orbital angular momentum; $< {l_p}0\,\,L\,\,0|\,{l_f}0 >$ is the Clebsch-Gordan coefficient and $U(L{l_f}{J_i}I;\,\,{l_p}{J_f})$ is the normalized Racah coefficient, 
$J_{i}$ is the total angular momentum of the system $p + A$ in the initial state of the radiative capture 
process, which is equal to the resonance spin, $J_{i}=J_{R}$, $I$ is the channel spin. The non-resonant capture amplitude describing the external direct capture 
in the $R$-matrix method is given by 
\begin{align}
 {M_{DC}} = 2C\,\sqrt {\frac{{{e^2}}}{{\hbar c}}{\lambda _N}\frac{{931.5}}{E}} \,{({k_\gamma }\,{r_p})^{L\, + 1/2}}\mu _{pA}^L[\frac{{{Z_p}}}{{m_p^L}} + {( - 1)^L}\frac{{{Z_A}}}{{m_A^L}}]\sqrt {\frac{{(L + 1){\hat L}}}{L}} \,\frac{1}{{{\hat L}!!}}\sqrt {{P_{{l_p}}}({k_{p,}}{r_p})}  \nonumber\\ 
 {W_{ - {\eta _f},{l_f} + 1/2}}({r_p})F({k_p},{r_p})\,G({k_p},{r_p}) < {l_p}0\,\,L\,\,0|\,{l_f}0 > U(L{l_f}{J_i}I;\,\,{l_p}{J_f})In{t_2}. 
\label{mdc1}
 \end{align}
As we can see the channel radiative width amplitudes and the direct capture amplitude are proportional to the same ANC because both describe the peripheral processes contributed by the tail of the overlap function whose amplitude is the ANC. Besides, the channel radiative width amplitude $\gamma_{\nu \,\gamma {\mathop{\rm (ext)}}}$  contains the proton reduced width amplitude $\gamma_{\nu \, p}$. Hence the relative sign of  $\gamma_{\nu \, \gamma {\mathop{\rm (ext)}}}$ and $M_{DC}$ depends on the sign of $\gamma_{\nu \, p}$.  It is also worth mentioning that
$\gamma_{\nu \, \gamma {\mathop{\rm (int)}} }$ is a fitting parameter and  
$\gamma_{\nu \, \gamma {\mathop{\rm (ext)}}}$ is a complex quantity because it contains an imaginary part. 
The radiative width is given by equation
\begin{align}
\Gamma_{\nu \gamma}= |\Gamma _{\nu \, \gamma {\mathop{\rm (int)}} }^{1/2} - \Gamma_{\nu \,\gamma {\mathop{\rm (ext)}}}^{1/2}|^{2} = 2\,k_{\gamma}^{3}\,|\gamma _{\nu \, \gamma {\mathop{\rm (int)}} } - \gamma_{\nu \,\gamma {\mathop{\rm (ext)}}}|^{2}
\label{Gammagamma1}
\end{align}
calculated at the $\nu$-th resonance energy. 
Another important point to underscore is that the signs of $\Gamma _{\nu \,\gamma {\mathop{\rm ext}}}^{1/2}$ and $M_{DC}$ relative to $\Gamma _{\nu\, \gamma {\mathop{\rm (int)}} }^{1/2}$ are synchronized. In all the fits presented below we use positive sign because it gave a better  fit. 

Finally the background resonance amplitude can be written as
\begin{equation}
M_{(BG)} = \frac{{\Gamma _\gamma ^{1/2}\Gamma _p^{1/2}}}{{E - {E_{{R_{(BG)}}}} + i\frac{{{\Gamma _{(BG)}}}}{2}}},
\label{mbgr1}
\end{equation}
where $E_{R_{(BG)}}$ is the resonance energy of the background resonance,
\begin{equation}
\Gamma _{\gamma (BG)}^{1/2}= \sqrt{2\,P_{l_{p}}(k_{p},r_{p})}\gamma_{\gamma (BG)},
\label{gammabgr1}
\end{equation}
\begin{equation}
\Gamma _{c(BG)}^{1/2}= \sqrt{2\,P_{l_{c}}(k_{c},r_{c})}\gamma_{c(BG)},
\label{gammabgr1}
\end{equation}
and $\gamma_{\gamma (BG)}$ is a complex radiative width amplitude for the decay of the background resonance to a bound state and $\gamma_{c (BG)}$ is the reduced width amplitude of the background resonance for the channel $c$..
The total resonance width of the background resonance is given by $\Gamma_{(BG)}= \Gamma_{p (BG) } + \Gamma_{ \alpha (BG)}$, where $\Gamma_{c (BG)}$ is the partial resonance width in the channel $c$. 
Note that all the energies are in MeV but the astrophysical factor $S(E)$ is in keVb. The dimension of the reduced particle width amplitude $\gamma_{\nu c}$ is MeV${}^{1/2}$ but $\gamma_{\nu \, \gamma}$ has dimension
MeV${}^{1/2}$fm${}^{3/2}$ for $L=1$. 

\section{Analysis}
 Altogether we performed 4 different sets of the fits, each set consists of two fits with the boundary conditions
$B_{c} = S_{c}(E_{2})$ and $B_{c}(E_{1})$. First we present two fits without a background pole. Then we perform 2 unconstrained fits called fits A(113), and 2 constrained fits called fits B(113).  In these fits we use all 113 data points of \cite{LeBlanc}. In addition, we performed also two fits A(70) with and without the background resonance using only 70 low-energy data points of \cite{LeBlanc} in the region of the first resonance.
For all the fits the channel radii in the proton and $\alpha$ channels, $r_{p}= 5.03$ fm and $r_{\alpha}= 7.0$ fm, and the ANC $C=14.154$ fm$^{-1/2}$  are kept the same as in \cite{muk08} and \cite{lacognata09} and the background resonance energy is fixed at $E_{R_{BG}}= 5.07$ MeV. All other parameters are varied to get a best fit.  For the case under consideration $l_{p}=0$, $\,l_{f}=1$, $J_{f}=0$, $\,J_{R}=1$ and $\,I=1$.

\begin{figure}
%[tbp]
\epsfig{file=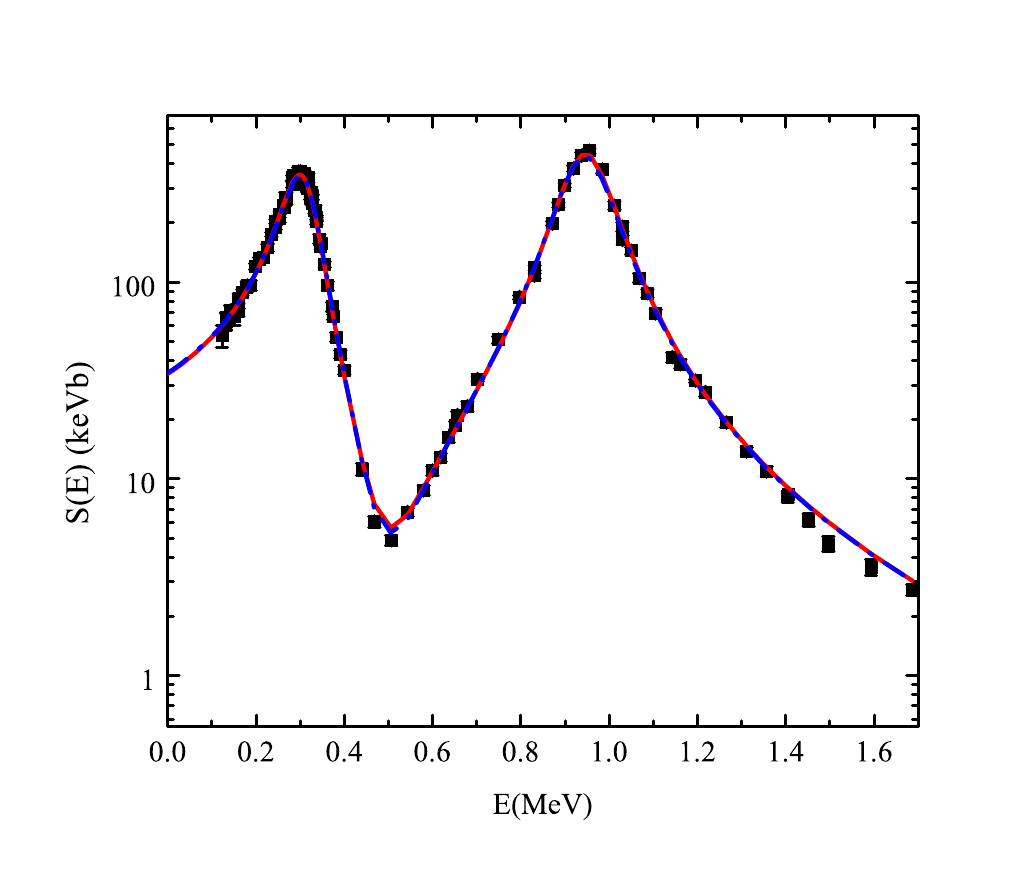,width=10cm}
\caption{(Color online) The astrophysical $S(E)$ factor for the $^{15}{\rm N}(p,\gamma)^{16}{\rm O}$  reaction. The black squares are experimental data from Ref. \cite{LeBlanc}. The red solid line is our unconstrained fit with the boundary condition $B_{c}= S_{c}(E_{2}=0.956\,{\rm MeV})$, which takes into account three interfering amplitudes: two $1^{-}$ resonances and non-resonant term. No background pole is included. The blue solid line is a similar fit with the boundary condition $B_{c}=S_{c}(E_{1}=0.30872\,{\rm MeV})$.
The ANC is $C=14.154$ fm${}^{-1/2}$ as in all other fits.}
\label{fig_noBGR}
\end{figure}

In Fig \ref{fig_noBGR} we demonstrate two best fits to 113 data points of \cite{LeBlanc} without any background pole with the boundary conditions in the channel $c=p,\,\alpha$ $\,\,B_{c}= S_{c}(E_{2})$ and $B_{c}= S_{c}(E_{1})$. The best fit with $\,\,B_{c}= S_{c}(E_{2})$ is achieved at $E_{2}=0.956$ MeV and $E_{1}=0.2872$ MeV and results in $S(0)=34.2$ keVb  with ${\tilde \chi}^{2}=2.6$ and total $\chi^{2}=273.1$. For the fit with $B_{c}=S_{c}(E_{1})$ at 
$E_{1}=0.30872$ MeV and $E_{2}= 1.0794$ MeV we get $S(0)= 34.6$ keVb and ${\tilde \chi}^{2}=2.5$  and $\chi^{2}= 259.9$.
Parameters of the fits are given in Tables \ref{table_1} and \ref{table_2}. As we can see both fits are quite good except for the bottom between two resonances and the high energy tail. From Fig. 9(b) \cite{LeBlanc} we can conclude that to get down ${\tilde  \chi}^{2}$ to the minimum one really needs to increase significantly the ANC as it has been done in \cite{LeBlanc}. 

\begin{figure}
%[tbp]
\epsfig{file=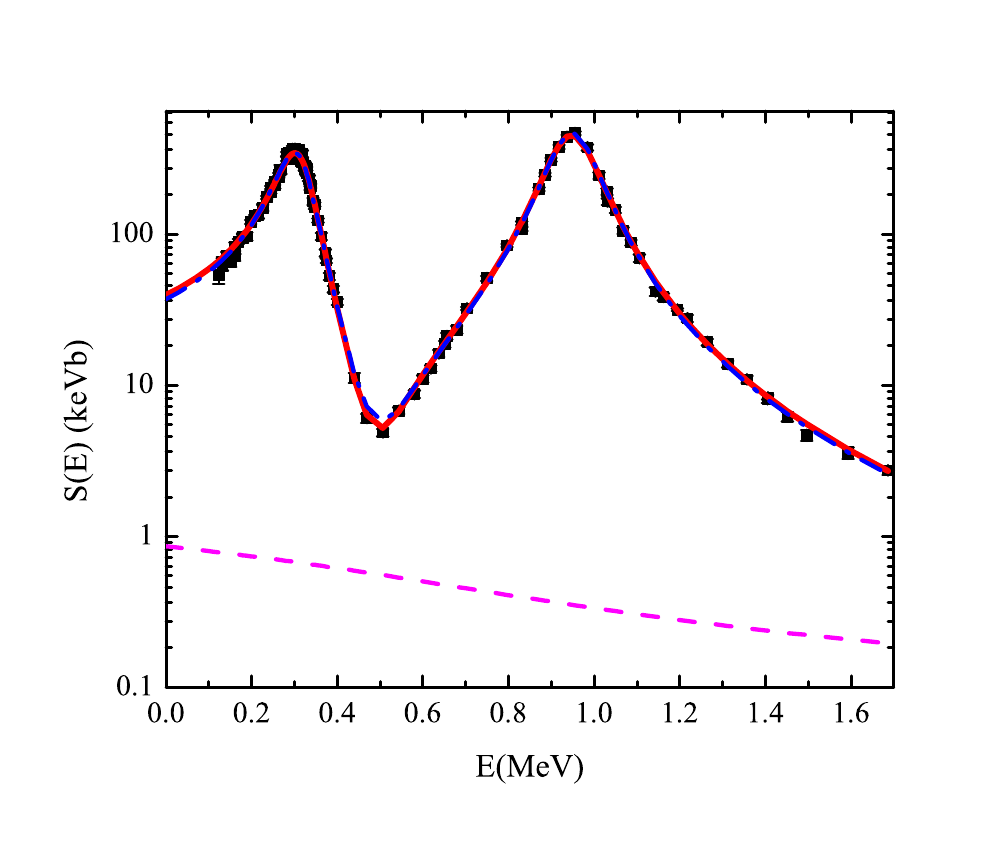,width=10cm}
\caption{(Color online) The astrophysical $S(E)$ factor for the $^{15}{\rm N}(p,\gamma)^{16}{\rm O}$  reaction. The black squares are experimental data from Ref. \cite{LeBlanc}. The red solid line is our unconstrained $R$-matrix fit A(113) with the boundary condition $B_{c}=S_{c}(E_{2})$, which takes into account four interfering amplitudes: two $1^{-}$ resonances, non-resonant term and background resonance at $5.07$ MeV. 
The blue dotted-dashed line is our unconstrained $R$-matrix fit A(113) similar to the previous one but with the boundary condition $B_{c}=S_{c}(E_{1})$.  The magenta solid line is the non-resonant $S(E)$ factor for the square of the ANC $C=14.154$ fm${}^{-1/2}$.}
\label{fig_SfactorAfit}
\end{figure}

Instead of varying the ANC to decrease ${\tilde \chi}^{2}$  we perform two unconstrained fits A(113) to the Notre Dame-LUNA data \cite{LeBlanc} adding a background resonance. All the parameters except for the ANC, channel radii and the background resonance energy are allowed to vary. First we have searched for the best fit for the boundary condition $\,B_{c}=S_{c}(E_{2})$,  $E_{2}$ is the energy of the second level which is taken close to the second resonance energy $E_{R_{2}}=0.9594$ MeV adopted in \cite{LeBlanc} while the first level is varied to get the best fit. We find the best fit at $E_{2}= 0.956$ MeV and $E_{1}=0.1662$ MeV. From this fit, using the Barker's transformation \cite{barker72,azuma}, we determine the $R$-matrix formal reduced widths  for the second resonance at resonance energy $E_{R_{2}}=0.9594$ MeV. After that we can find the observable partial resonance widths for the channel $c$ using the standard $R$-matrix equation \cite{barker08}
\begin{equation} 
{\tilde \Gamma}_{\nu \, c}= \frac{2\,\gamma_{\nu \, c}^{2}\,P_{c}(E_{{R_{\nu}}})}{1+ \sum\limits_{c' = p,\alpha } {\gamma _{\nu \, c'}^2\,\frac{{d{S_{c'}}}}{{dE}}{|_{E = {E_{{R_{\nu}}}}}}}},
\label{observwidths1}
\end{equation}
where $E_{R_{\nu}}$ is the resonance energy of the level $\nu$. 
In the second unconstrained fit A(113), we have searched for the best fit with the boundary condition $\,B_{c}=S_{c}(E_{1})$, where the energy of the first level $E_{1}$ is near the first resonance at $E_{R_{1}}=0.3104$ MeV adopted in \cite{LeBlanc} while the second level is varied to get the best fit.  
 For this boundary condition we find the best fit at $E_{1}=0.30872$ MeV and $E_{2}= 1.0576$ MeV. The formal reduced widths and observable resonance widths for the first resonance are determined from this fit by shifting the boundary condition from $E_{1}=0.30872$ MeV to the first resonance location at $E_{R_{1}}=0.3104$ MeV and using Eq. (\ref{observwidths1}).  
In Fig. \ref{fig_SfactorAfit} we demonstrate the astrophysical factors $S(E)$ obtained from these two unconstrained A(113) with fixed ANC $C=14.154$ fm${}^{-1/2}$ and the background resonance included.
The red solid line is the fit corresponding to the boundary condition at $E_{2}= 0.956$ MeV with the normalized ${\tilde \chi}^{2}= 1.84$. This fit is practically identical to the one in \cite{LeBlanc} resulting in $S(0)=39.0$ keVb in agreement with \cite{LeBlanc}. For the fit A(113) with the boundary condition at $E_{1}=0.30872$, the blue dotted-dashed line, we obtain ${\tilde \chi}^{2}= 1.76$  with $S(0)=37.2$ keVb. This fit goes slightly lower than the red line at low energies better reproducing the low-energy trend of the data.  
The magenta solid line represents the non-resonant $S(E)$ factor for the ANC $C=14.154$ fm${}^{-1/2}$ which has been used for both fits. This ANC is within the experimental interval $13.86 \pm 0.91$ fm$^{-1/2}$ determined from the  ${}^{15}{\rm N}({}^{3}{\rm He},d){}^{16}{\rm O}$ reaction \cite{muk08}. Thus adopting a physical ANC we correctly fix the normalization of the external direct capture amplitude and the channel radiative width amplitude, and adding the background pole rather than varying the ANC way beyond experimental limits \cite{LeBlanc} we are able to get the same fit as in \cite{LeBlanc}.
The obtained formal reduced widths and observable resonance widths for the fits A(113) along with the corresponding parameters from \cite{LeBlanc} and \cite{barker2008} are given in Tables \ref{table_1} and \ref{table_2}. 
\begin{table}
\begin{center}
\caption{Resonance parameters. Parameters of the $R$-matrix fits to the $^{15}{\rm N}(p,\gamma)^{16}{\rm O}$
capture 113 data points \cite{LeBlanc} along with the fitting parameters from \cite{LeBlanc} and \cite{barker2008}. Fits A(113) and B(113) are our fits. The fitting parameters for the second resonance are determined from adopting the boundary condition $B_{c}=S_{c}(E_{2})$ at the second energy level $E_{2}=0.956$ MeV, which is near the second resonance $E_{R_{2}}=0.9594$ MeV, with the first energy level $E_{1}=0.1662$ MeV found from the fit. The channel radii, $r_{p}=5.03$ fm and $r_{\alpha}=7.0$ fm,  and ANC $C=14.154$ fm${}^{-1/2}$ have been used in all the fittings.  
In all the fittings where the background resonance is included the background resonance energy is $5.07$ MeV,
the proton reduced width amplitude of the background resonance is $\gamma_{p(BG)}= -0.3$ MeV${}^{1/2}$ and the $\alpha$ reduced width amplitude $\gamma_{\alpha(BG)}= 0.07$ MeV${}^{1/2}$. In the fit A(113) the search for the best fitting has been performed using unconstrained variation of other parameters. Using the Barker's transformation \cite{barker72,azuma} the fitting parameters are transformed to the ones corresponding to the boundary condition at the second resonance energy $E_{R_{2}}=0.9594$ MeV adopted in \cite{LeBlanc}. These parameters are given as the fitting parameters for the second resonance. The radiative width for the background pole is found to be $\Gamma_{\gamma (BG)}=354.9$ eV. To determine the fitting parameters for the first resonance the boundary condition $B_{c}=S_{c}(E_{1})$ has been adopted, where the first energy level $E_{1}=0.30872$ MeV. The second energy level found from the fit is $E_{2}=  1.0576$ MeV. The fitting parameters are transformed to the ones corresponding to the boundary condition at the first resonance $E_{R_{1}}= 0.3104$ MeV adopted in \cite{LeBlanc} and are shown in the table as the fitting parameters for the first resonance. The radiative width for the background pole is found to be $\Gamma_{\gamma(BG)}=360.5$ eV. In the constrained fits B(113) the procedure is the same as described before but two more parameters are fixed. When searching for the fitting parameters for the $\nu$-th resonance the proton and $\alpha$ reduced width amplitudes of the resonance $\nu$ are fixed. Their values are taken from the ${}^{15}{\rm N}(p,\alpha){}^{12}{\rm C}$ data fitting parameters \cite{barker08,lacognata09}.
In the first constrained fit B(113), which is used to determine the parameters for the second resonance, the boundary condition $B_{c}=S_{c}(E_{2})$ is taken at $E_{2}= 0.956$ MeV with the first energy level found from the fit  $E_{1}=0.1697$ MeV. The radiative width for the background pole is found to be $\Gamma_{\gamma(BG)}=129.3$ eV. To determine the parameters for the first resonance we use the boundary condition $B_{c}=S_{c}(E_{1})$ near the first resonance $E_{1}=0.30872$ MeV and found from the fit the second energy level $E_{2}=1.0576$ MeV. The rest is the same as in fit A(113). The radiative width for the background resonance is $\Gamma_{\gamma(BG)}=283.1$ eV. $\,{\tilde \Gamma}_{\nu\,c}$ is the observable resonance partial width in the channel $c$ and $\Gamma_{\nu\,\gamma}$ is the $\nu$-th resonance radiative width.
\label{table_1}}
%\vspace{0.4cm}
\begin{tabular}{ccccccccccccc}
%[t]
\hline \hline \\
Reference & $\gamma_{1\, p}^{2}$ [keV]  & ${\tilde \Gamma}_{1\,p}$ [keV]&  $\gamma_{1\, \alpha}^{2}$ [keV] & 
${\tilde \Gamma}_{1\, \alpha}$ [keV] &   $\Gamma_{1\, \gamma }$  [eV] &  $\gamma_{2\, p}^{2}$ [keV] &  
${\tilde \Gamma}_{2\,p}$[keV]& $\gamma_{2\, \alpha}^{2}$ [keV] &  ${\tilde \Gamma}_{2\, \alpha}$ [keV] &  $\Gamma_{2\, \gamma}$ [eV]   \\
\hline\\[1mm]
%\cite{rolfs74}   & $1.1$ & $100$   &  $98$  & $45$ &  $12 \pm 2$ &  $32 \pm 5$  \\
\protect \cite{LeBlanc}  & $52.8$ &  $0.20$   & $13.5$ & $112.0$ &  $33.8$ &   $309.1 $ & $110.6$ &  $5.0$ &  $40.6$   & $38.7$   \\
\hline \\
\cite{barker08},\\ 
Table II, HH(c)  & $355.2$ & $1.0$ & $10.6$ &  $85.8$ &      &  $265.2$  & $98$ & $5.4$ & $40.3$ &     \\     
\hline\\
present work, unconstrained fits\\ 
A(113)  & $358.8$ & $1.3$  & $14.4$ & $99.4$ &  $7.5 $ &    $221.6$ & $82.8$ & $ 7.5$ & $63.2$  & $63.6$   \\
\hline\\
present work, constrained fits\\
B(113)  & $259.8$ & $1.0$  & $13.6$ & $99.7$ &  $9.3 $  &  $  268.9$ & $98.1$ & $ 6.0$ & $49.4$  & $54.4$   \\  
\hline \\
present work, unconstrained fits \\
without background resonance  & $353.3$ & $1.3$  & $14.1$ & $98.3$ &  $7.5 $  &  $231.4$ & $86.0$ & $ 6.9$ & $58.2$  & $57.6$ \\
[2mm]
%\vspace{0.4cm}
\hline \hline \\
\end{tabular}
\end{center}
\end{table}
%\vspace{0.4cm}

\begin{table}
\begin{center}
\caption{\,Internal and external radiative width amplitudes for the first and second resonances. The 
amplitudes for the second resonance at $E_{R_{2}}= 0.9594$ MeV (first resonance at $E_{R_{1}}=0.3104$ MeV) are determined from the unconstrained fit A(113) and constrained fit B(113) for the boundary condition at the second (first) resonance. \label{table_2}} 
\begin{tabular}{ccccc}
%[t]
\hline \hline \\
Fits & $\gamma_{1\, \gamma (int)}$ [MeV$^{1/2}$ fm$^{3/2}$] \,\, &\,\, $\gamma_{1\,\gamma (ext)}$ [MeV$^{1/2}$ fm$^{3/2}$]\,\, &\,\,  $\gamma_{2\,\gamma (int)}$ [MeV$^{1/2}$ fm$^{3/2}$]\,\, & \,\, $\gamma_{2\,\gamma (ext)}$ [MeV$^{1/2}$ fm$^{3/2}$]  \\
\hline\\[1mm]\\ 
\cite{LeBlanc}           &  $0.22$   &                  &    $0.19$       &    \\
\hline\\
\cite{barker08}         &   $0.085$  &                   &   $0.24$       &     \\  
\hline\\
Unconstrained fits A(113)  & $ 0.062$ &  $ 0.061 + i\,0.000059$ &    $0.28$ &   $ 0.053 + i\,0.0041$   \\
\hline\\
Constrained fits B(113)  & $ 0.085$ &  $ 0.052 + i\,0.000054 $  &  $0.25$ &  $ 0.059 + i\,0.0044 $   
\\   
\hline\\
Unconstrained fits \\
without background resonance   & $0.062$ &  $0.060 + i\,0.000059 $  &  $ 0.26$ &  $0.055 + i\,0.0042 $
\\
[2mm]
%\vspace{0.4cm}
\hline \hline \\
\end{tabular}
\end{center}
\end{table}

\begin{figure}
%[tbp]
\epsfig{file=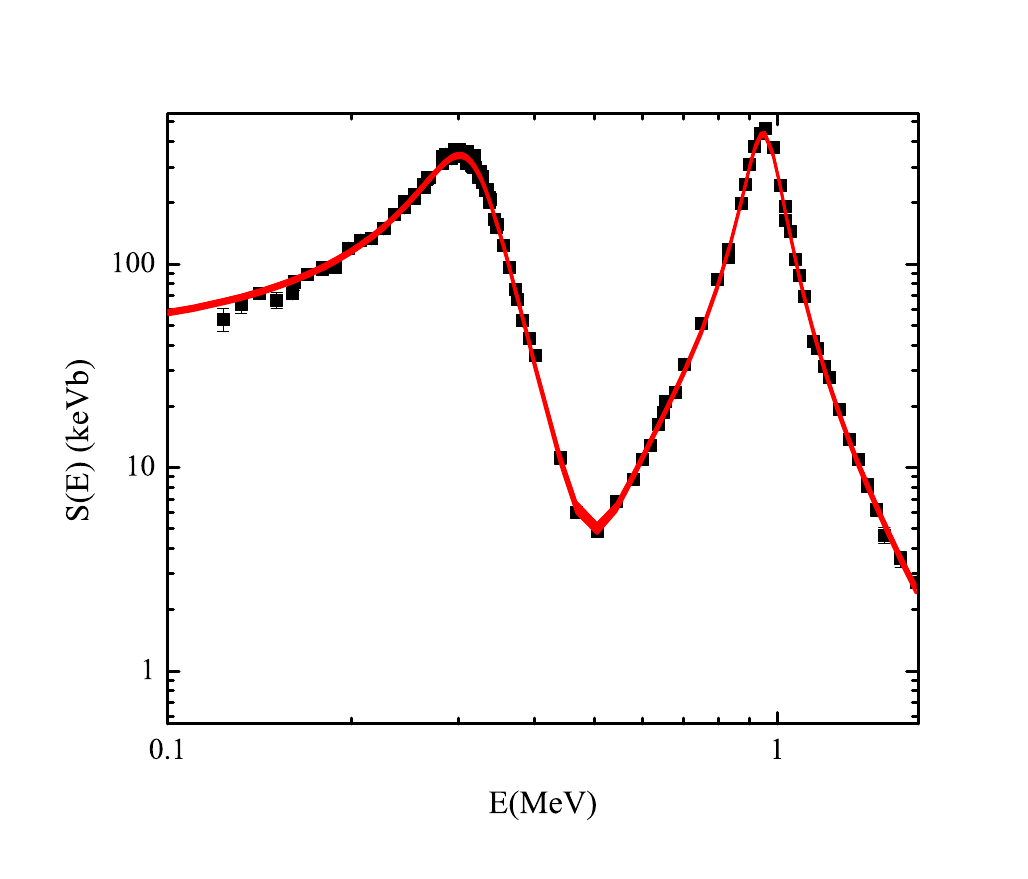,width=10cm}
\caption{(Color online) The astrophysical $S(E)$ factor for the $^{15}{\rm N}(p,\gamma)^{16}{\rm O}$.
The band for the astrophysical factor $S(E)$ obtained from  the unconstrained fit A(113) with the boundary condition $B_{c}=S_{c}(E_{2})$. 
The upper and low limits of the band correspond to the fitting of the experimental data, which deviate by  
$1\,\sigma$ up and down from the center, correspondingly. Note that the borders of the band have practically the same particle reduced width amplitudes $\gamma_{\nu\,c}$ for the first and second levels.
The proton partial width for the second resonance within the band is ${\tilde \Gamma}_{2\,p}= 82.8 \pm 0.6$ keV,
the $\alpha$-particle partial width for the second resonance ${\tilde \Gamma}_{2\,\alpha}= 63.2 \pm 0.9$ keV,  the radiative width of the second resonance $\Gamma_{{2}\,\gamma}= 63.6 \pm 2.4$ eV and the radiative width of the background pole $\Gamma_{\gamma(BG)}= 354.9 \pm 23.6$ eV. 
The black squares are experimental data from Ref. \cite{LeBlanc}.}
\label{fig_bandpaper}
\end{figure}

In Fig \ref{fig_bandpaper}, in the logarithmic scale for both axes, we show the band between upper and low limits of
the astrophysical factor $S(E)$ obtained from the unconstrained fit A(113) with the boundary condition $B_{c}=S_{c}(E_{2})$ at $E_{2}=0.956$ MeV for the energy region $E < 1.7$ MeV. The upper (lower) limit with $S(0)=40.1$ keVb ($S(0)=37.9$ keVb) and ${\tilde \chi}^{2}= 3.0$ (${\tilde \chi}^{2}=2.6$) of the band corresponds to the fitting of the experimental data which deviate by $1\,\sigma$ up (down) from the center which corresponds to $S(0)= 39.0$ keVb with ${\tilde \chi}^{2}= 1.84$.
Thus, taking into account the experimental uncertainties given in \cite {LeBlanc}, we can conclude that our unconstrained fit A(113) with the boundary condition $B_{c}=S_{c}(E_{2})$ results in $S(0)= 39.0 \pm 1.1$ keVb.
However, the logarithmic scales for both axes show the problem with the fitting at low energies, where the fit A(113) with the boundary condition $B_{c}=S_{c}(E_{2})$ deviates from the experimental trend. A similar trend is present in the fit of \cite{LeBlanc}.
The reason for this trend is that our fits and fit in \cite{LeBlanc} have been performed minimizing the weighted $\chi^{2}$ with the weights $\Delta_{i}^{-2}$, where $\Delta_{i}$ is the experimental uncertainty at point $i$. Since the relative experimental uncertainties at low energies are larger than in the region between the two resonances and at higher energies, the weighted fit underestimates the importance of the low energy region, which is the most crucial for determination of the $S(0)$ astrophysical factor.

We can compare the results of the fits A(113) with \cite{LeBlanc} and \cite{barker08}. Note that from different fits presented in \cite{barker08} we choose the fit HH(c), Tables II and III, to the data of \cite{hebb60}, because they are pretty close to the data of \cite{LeBlanc} in the region of the first resonance  and the selected fit from \cite{barker08} resulted in $S(0)=35.2$ keVb, which agrees with our results and close to \cite{LeBlanc} $S(0)= 39.6 \pm 2.6$ keVb.

 As it has been mentioned in \cite{barker08}, there is significant uncertainty in the values of the proton and $\alpha$ partial widths for the second resonance
what can be also concluded from compilation \cite{TUNL}.  That is why there is no recommended values for these widths in \cite{barker08} after analysis of the ${}^{15}{\rm N}(p,\alpha){}^{12}{\rm C}$ data. One of the problems is that the ${}^{15}{\rm N}(p,\alpha){}^{12}{\rm C}$ and ${}^{15}{\rm N}(p, \gamma){}^{16}{\rm O}$ 
reactions put a limitation on the ratio $\gamma_{1 \alpha}^{2}/\gamma_{2 \alpha}^{2}$  and the $E1$ strength ratio of the second and first resonances, which should be equal due to the isospin mixture of two $1^{-}$ resonances.
This isospin mixture can be written as \cite{hebb60,rolfs74,barker08}
\begin{align}
\psi_{1}= \alpha\,|T=0> + \beta\,|T=1>, \nonumber\\
\psi_{2}= \beta\,|T=0> - \alpha\,|T=1>, 
\label{isospinmixture1}
\end{align}
where $\alpha^{2} + \beta^{2}=1$.
Since the $\alpha$-particle decays of these resonance in ${}^{16}{\rm O}$ to the ground state are allowed  
only due to the $T=0$ components, we have
\begin{align}
\frac{ \gamma_{1\,\alpha}^{2} }{ \gamma_{2\,\alpha}^{2} }= \frac{\alpha^{2}}{\beta^{2}}.
\label{aredwidtratio1}
\end{align}
Correspondingly, the strength of the $E1$ decays of these resonances to the ground state is entirely determined by the $T=1$, i.e. 
\begin{align}
\frac{E1_{2}}{E1_{1}}=\frac{\alpha^{2}}{\beta^{2}}.
\label{gE1strength1}
\end{align}
 From the fit in \cite{LeBlanc} one gets the ratio $\gamma_{1\,\alpha}^{2}/\gamma_{2 \alpha}^{2}= 2.7$ and 
$\gamma_{1\,\alpha}^{2}/\gamma_{2 \alpha}^{2}= 1.96$  from \cite{barker08}.
To get the ratio of the $E1$ intensities we remind that $\gamma_{\nu\, \gamma (int)} \sim <\varphi|{\hat O}|\psi_{\nu (int)}>|_{r \leq r_{p}}$, where $\varphi$ is the $({}^{15}{\rm N}\,p)$ bound state wave function, ${\hat O}$ is the electromagnetic operator and $\psi_{\nu (int)}$ is the internal resonant wave function of the $\nu$-th resonance given by a standing wave satisfying Eq. (\ref{isospinmixture1}). Then the ratio of the $E1$ intensities can be estimated from the ratio of $\gamma_{\nu\, \gamma (int)}^{2}$ assuming the dominance of the internal contribution to the electromagnetic transition matrix element. From \cite{LeBlanc} we get  $\gamma_{2\, \gamma (int)}^{2}/\gamma_{1\, \gamma (int)}^{2}=0.66$ and $\gamma_{2\, \gamma (int)}^{2}/\gamma_{1\, \gamma (int)}^{2}=7.97$ from \cite{barker08}. If we use for the $E1$ intensity ratio of the total radiative widths we get from \cite{LeBlanc} $E1_{2}/E1_{1}=(\Gamma_{2\,\gamma}/k_{2\,\gamma}^{3})/(\Gamma_{1\,\gamma}/k_{1\,\gamma}^{3})=0.98$, where $k_{\nu\,\gamma}= (E_{R_{\nu}}+ \varepsilon)/{\hbar\,c}$ is the momentum of the emitted photon for transition from the resonance $\nu$ to the ground state with the proton binding energy 
$\varepsilon$. Thus the ratio of the $\alpha$ reduced widths from \cite{LeBlanc} is pretty consistent with findings in \cite{barker08,lacognata09} but deviates from \cite{barker2008}, while the ratio of the radiative resonance widths is too small compared to all the previous estimations due to too high radiative width of the first resonance which was estimated to be around $10$ eV \cite{hebb60,rolfs82,TUNL}. The proton partial width ${\tilde \Gamma}_{2p}= 110$ keV is higher than the previous estimations \cite{barker08,barker2008,TUNL}. Note that we do not include estimations from the analysis of the data of \cite{rolfs74}.

The partial widths for the first resonance are better known than for the second one. According to \cite{barker08} and \cite{barker2008} ${\tilde \Gamma}_{1p}=1.1$ keV and $1.0$ keV, correspondingly, and ${\tilde \Gamma}_{1\, \alpha}= 92 \pm 8$ keV and different previous estimations are pretty close to these values \cite{TUNL}. Note that all the widths are in the center-of-mass system. That is why too
low value of ${\tilde \Gamma}_{1 p}= 0.2$ keV obtained in \cite{LeBlanc} is difficult to explain.   
Our unconstrained fits A(113) is not satisfactory although it better agrees with the previous estimations for the first resonance than the fit of \cite{LeBlanc}. The quoted value $\Gamma_{1 \gamma}= 12 \pm 2$ eV in \cite{TUNL} was taken from \cite{rolfs74}, while \cite{hebb60} obtained from a single level analysis (only the first resonance was included) $\Gamma_{1 \gamma}= 8$ eV, and from the two-level analysis $\Gamma_{1 \gamma}= 12.8$ eV. 
Our $\Gamma_{1 \gamma}=7.5$ eV is significantly lower than the corresponding value in \cite{LeBlanc} 
and closer to \cite{hebb60,rolfs82}. Our $\Gamma_{2\,\gamma}=63.6$ eV is higher than $\Gamma_{2\, \gamma}=
32 \pm 5$ eV \cite{rolfs82}, $\,38.7$ eV \cite{LeBlanc} and $44 \pm 8$ eV obtained from the branching ratio \cite{TUNL} but lower than $88$ eV \cite{hebb60}. Our ${\tilde \Gamma}_{1 \alpha}=99.4$ keV is in a perfect aggreement with estimation $92 \pm 8$ keV \cite{barker08} and with other estimations \cite{TUNL} while the value ${\tilde \Gamma}_{1 \alpha}= 112.0$ keV \cite{LeBlanc} looks beyond of the boundaries of the existing estimations.  However, our  $\gamma_{1\, \alpha}^{2}/\gamma_{2\, \alpha}^{2}= 1.92$ is much smaller than  $(\Gamma_{2\,\gamma}/k_{2\,\gamma}^{3})/(\Gamma_{1\,\gamma}/k_{1\,\gamma}^{3})=7.2$.

\begin{figure}
%[tbp]
\epsfig{file=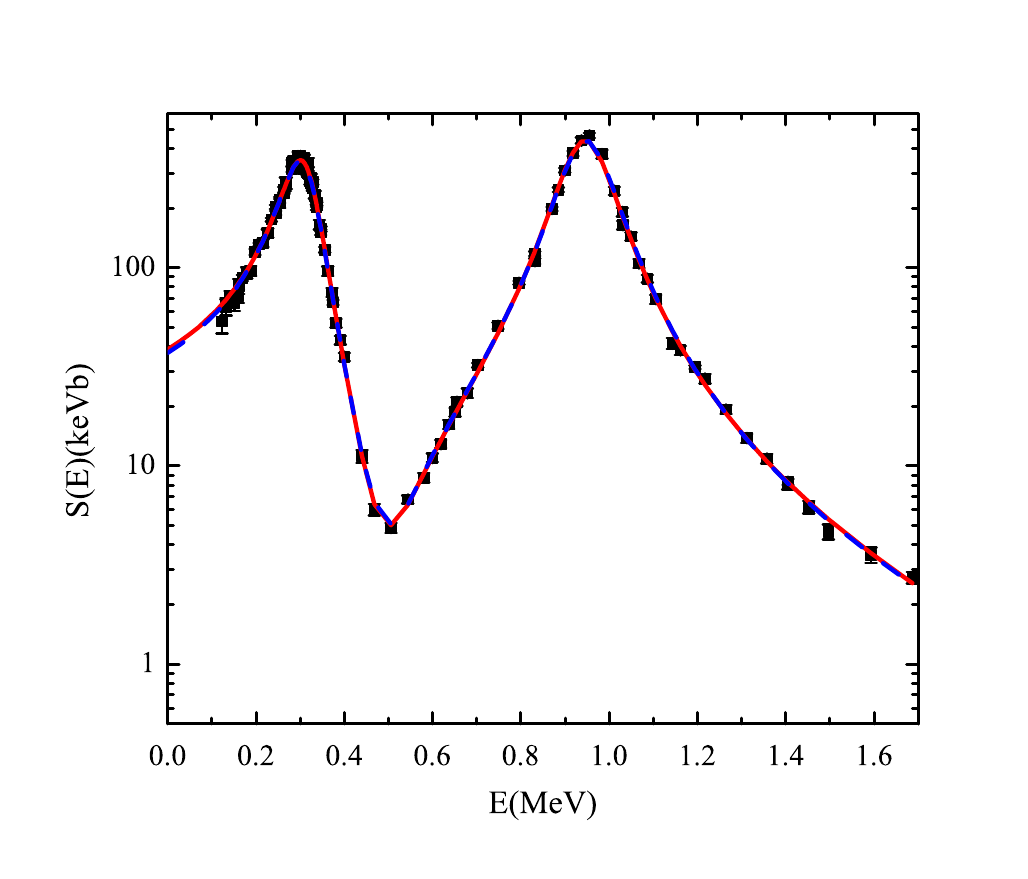,width=10cm}
\caption{(Color online) The astrophysical $S(E)$ factor for the $^{15}{\rm N}(p,\gamma)^{16}{\rm O}$.
The red solid line is the constrained fit B(113) with the boundary condition $B_{c}=S_{c}(E_{2}=0.956\, {\rm MeV})$, the blue dotted-dashed line is the constrained fit B(113) with the boundary condition $B_{c}=S_{c}(E_{1}= 0.30872\, {\rm MeV})$.  The black squares are experimental data from Ref. \cite{LeBlanc}. The ANC is $C=14.154$ fm${}^{-1/2}$.}
\label{fig_B}
\end{figure}

Due to the above mentioned problems with the unconstrained fits A(113) we performed two constrained fits B(113) keeping in mind that with a slightly larger ${\tilde \chi}^{2}$ than for the unconstrained fits we can get more reasonable fitting parameters. Once again we did two different fits corresponding to two boundary conditions with parameters given in Tables \ref{table_1} and \ref{table_2}. In these fits, in addition to the fixed the channel radii in the proton and $\alpha$ channels, $r_{p}= 5.03$ fm and $r_{\alpha}= 7.0$ fm, the ANC $C=14.154$ fm$^{-1/2}$ and the background resonance energy $E_{R_{BG}}= 5.07$ MeV, we also fix $\gamma_{\nu\, c}$, $\,c=p,\,\alpha$, when the boundary condition is chosen near the resonance energy $E_{R_{\nu}}$. 
These reduced widths are taken from the analysis of the direct $(p,\alpha)$ data \cite{rolfs82,barker08,lacognata09} and indirect data \cite{lacognata09}. First we adopt the boundary condition near the second resonance at $E_{2}= 0.956$ MeV with the energy of the first level $E_{1}= 0.170$ MeV. For the best fit we get ${\tilde \chi}^{2}=1.93$ and $S(0)=38.8$ keVb. After that we set up the boundary condition at $E_{1}=0.30872$ MeV. 
For the best fit for the energy of the second level $E_{2}=1.0573$ MeV  we get ${\tilde \chi}^{2}=1.74$ \footnote[1]{We note that the total $\chi^{2}$ for the unconstrained fit A(113) with the boundary condition $B_{c}=S_{c}(E_{1})$ is slightly smaller than for the corresponding constrained fit B(113). However, because 
for the constrained fit the number of the fitting parameters are smaller than for the unconstrained one the normalized ${\tilde \chi}^{2}$ for the constrained fit is slightly smaller than for the unconstrained.}
with $S(0)=37.2$ keVb. The parameters given in Tables \ref{table_1} and \ref{table_2} are obtained for the boundary conditions at the resonance energies adopted in \cite{LeBlanc}. Barker's transformation to get the fitting parameters at the energy of the first resonance practically didn't change them because of the proximity of our
adopted first level $E_{1}=0.30872$ MeV and the first resonance energy $E_{R_{1}}=0.3104$ MeV adopted in \cite{LeBlanc}. In Fig. \ref{fig_B} the $S(E)$ factors are shown for both constrained fits B(113): the solid red line represents the fit with the boundary condition at $E_{2}= 0.956$ MeV and the blue dotted-dashed line, which better reproduces the low-energy experimental trend, corresponds to the fit with the boundary condition at $E_{1}=0.30872$ MeV. 
The constrained fits B(113) have parameters which better agree with previous estimations \cite{TUNL} than unconstrained fits A(113). In particular, ${\tilde \Gamma}_{2\,\alpha}$ is lower and $\Gamma_{1\,p}$  better agrees with previous estimations \cite{TUNL} than the width obtained in \cite{LeBlanc}. The same is true for the observable partial widths for the first resonance which agree with previous estimations and the radiative width for the first resonance is in better agreement with \cite{TUNL}. Our $\Gamma_{2\,\gamma}=54.4$ eV, although lower than in the fit A(113), is still high and remains the only problem to get consistency with previous estimations \cite{barker08,barker2008,lacognata09,TUNL}. Because of that the ratio   
$(\Gamma_{2\,\gamma}/k_{2\,\gamma}^{3})/(\Gamma_{1\,\gamma}/k_{1\,\gamma}^{3})=5.0$ is still too high compared to $\gamma_{1\, \alpha}^{2}/\gamma_{2\, \alpha}^{2}= 2.3$ which is fixed in agreement with ${}^{15}{\rm N}(p,\alpha){}^{12}{\rm C}$ \cite{barker08,lacognata09}. 
In Table \ref{table_3} we present ${\tilde \chi}^{2}$ and $S(0)$ astrophysical factors for all our fits and fit from \cite{LeBlanc}. The uncertainties of our $S(0)$ factors are obtained by fitting to the upper (lower) border of the data obtained by adding (subtractiing) the experimental uncertainties to the experimental astrophysical factor at each point. A similar procedure has been used to determine the band shown in Fig. \ref{fig_bandpaper}. 
\begin{table}
\begin{center}
\caption{$\,{\tilde \chi}^{2}$ and $S(0)$ astrophysical factors for the ${}^{15}{\rm N}(p,\,\gamma){}^{16}{\rm O}$ capture process obtained from our fits and from the fit in \cite{LeBlanc}.
\label{table_3}} 
\begin{tabular}{ccc}
%[t]
\hline \hline \\
Fits & \,\, ${\tilde \chi}^{2}$\, \,&\,\, $S(0)$ [keVb] \\
\hline\\[1mm]\\ 
Ref. \cite{LeBlanc} \,\,   & \,\,$ 1.80$ & \,\, $39.6 \pm 2.6  $\\
\hline \\
Fit A(113), $\,B_{c}= S_{c}(E_{2})$\,\,&\,\, $1.84$ &\,\,  $39.0 \pm 1.1$ \\
\hline\\
Fit A(113),$\,B_{c}= S_{c}(E_{1})$ \,\,  & \,\, $ 1.76$ &  $37.2 \pm 1.0$ \\     
\hline\\
Fit B(113),$\,B_{c}= S_{c}(E_{2})$  \,\,   &\,\, $ 1.93$ &  $38.8 \pm 1.1$\\
\hline\\
Fit B(113), $\,B_{c}= S_{c}(E_{1})$ \,\,    &\,\, $ 1.74$ &  $37.2 \pm 1.0$\\
\hline\\
Fit without background resonance, \\
$\,B_{c}= S_{c}(E_{2})$                \,\,&\,\, $2.58$ &\,\,  $34.1 \pm 1.0$ \\
\hline \\
Fit without background resonance, \\   
$\,B_{c}= S_{c}(E_{1})$           & \,\,$2.45$ & \,\, $34.6 \pm 1.0  $\\
[2mm]
%\vspace{0.4cm}
\hline \hline \\
\end{tabular}
\end{center}
\end{table}
As we can see from Table \ref{table_3} five different fits result in quite stable $S(0)$ factors ranging 
in the interval $33.1 \leq S(0) \leq 40.1$ keVb.
The unconstrained fit A(113) and constrained fit B(113) with $B_{c}=S_{c}(E_{1})$  give the minimum ${\tilde \chi}^{2}$ among all our fits with $S(0)=37.2 \pm 1.0$ keVb.  which overlaps with the result reported in \cite{LeBlanc}. However, as we have discussed, the constrained fits B(113) yield fitting parameters including a correct ANC value, which are more consistent with the previous estimations. Besides, our both best fits better reproduce the low-energy slope of the $S(E)$ astrophysical factor than the unconstrained fit A(113) with $B_{c}=S_{c}(E_{2})$, see Fig \ref{fig_bandpaper}, and the fit of \cite{LeBlanc}, which trend away from the low energy data, and a lower value of the $S(0)$ is quite plausible when lower energy data will be available.
We included also into the list of the fits two fits performed without a background resonance because their ${\tilde \chi}^{2}$ deviate from the minimum ${\tilde \chi}^{2}$ by $< 1$. These fits provide the lowest $S(0)$
better reproducing the low energy behavior of the $S(E)$ factor than the ones with higher $S(0)$. 

\begin{figure}
%[tbp]
\epsfig{file=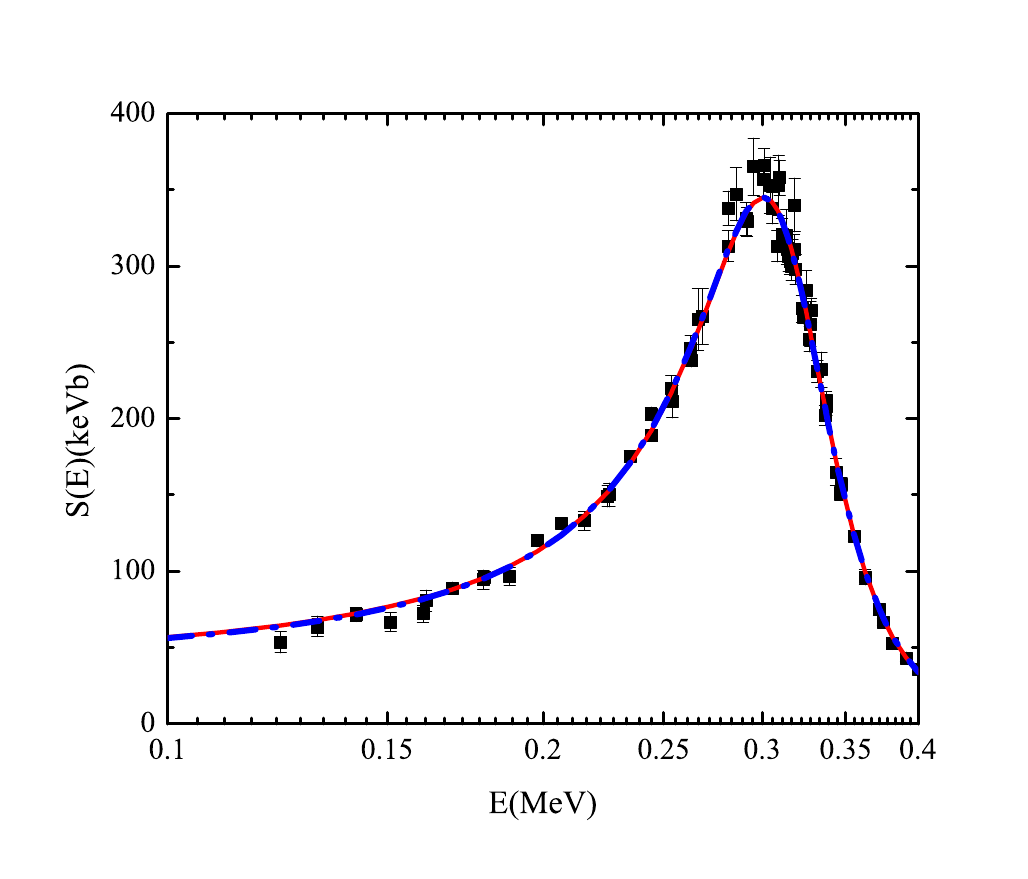,width=10cm}
\caption{(Color online) The astrophysical S(E) factor for the $^{15}{\rm N}(p,\gamma)^{16}{\rm O}$  reaction in the low-energy region including the first resonance (70 data points). The black squares are the experimental data from Ref. \cite{LeBlanc}. The red solid line is the unconstrained fit A(70) for the boundary condition $B_{c}=S_{c}(E_{1}=0.30872\, {\rm MeV})$;  for this fit $E_{2}=1.056$ MeV,  $\gamma_{2\,\gamma(int)}=0.056$ MeVfm$^{3/2}$, $\,\gamma_{2\,\gamma(ext)}= 0.0607 + i\,0.000059$ MeVfm$^{3/2}$. The blue dotted-dashed line is the constrained fit B(70) with the boundary condition $B_{c}=S_{c}(E_{1}= 0.30872\, {\rm MeV})$; for this fit $E_{2}=1.05$ MeV, $\gamma_{2\,\gamma(int)}=0.082$ MeVfm$^{3/2}$, $\,\gamma_{2\,\gamma(ext)}= 0.052 + i\,0.000051$ MeVfm$^{3/2}$. The ANC is $C=14.154$ fm${}^{-1/2}$.}
\label{fig_Firstpeak}
\end{figure}

To increase the weight of the low-energy points we present also two fits to 70 low-energy data points in the region of the first resonance rather than to all 113 data points: the unconstrained fit A(70) and the unconstrained fit B(70)  with the boundary condition $B_{c}=S_{c}(E_{1})$.  New fits to 70 data points are shown in Fig. \ref{fig_Firstpeak} and parameters are given in Tables \ref{table_Firstpeak}  and \ref{table_chiS70}. We use the the logarithmic scale for the energy axis and linear scale for the $S(E)$ factor to see more clearly the low-energy behavior of the astrophysical factors. 

It is worth mentioning that both fits to 70 data points favor lower value of the $S(0)$ factors than in 
\cite{LeBlanc}, i.e. the same tendency which we have observed for the corresponding fits to 113 data points. 
The parameters of both fits are similar to the ones of the fits for 113 data points.
\begin{table}
\begin{center}
\caption{Resonance parameters. Parameters of the unconstrained A(70) and constrained B(70) fits to the $^{15}{\rm N}(p,\gamma)^{16}{\rm O}$ 70 data points \cite{LeBlanc}. Physical meaning of the parameters and the procedure are similar to the one described in the caption for Table \ref{table_1}. Note that for both fits  A(70) and B(70) we present the parameters only for the first resonance  because the boundary condition is adopted at the energy of the first level.
\label{table_Firstpeak}}
%\vspace{0.4cm}
\begin{tabular}{cccccc}
%[t]
\hline \hline \\
Fit & $\gamma_{1\, p}^{2}$ [keV]  & ${\tilde \Gamma}_{1\,p}$ [keV]&  $\gamma_{1\, \alpha}^{2}$ [keV] & 
${\tilde \Gamma}_{1\, \alpha}$ [keV] &   $\Gamma_{1\, \gamma }$  [eV]  \\
\hline\\[1mm] 
fit A(70), \\
$B_{c}=S_{c}(E_{1 }=0.30872\, {\rm MeV})$ & $358.9$ & $1.4$  & $14.3$ & $98.9$ &  $6.9$      \\
\hline\\
fit B(70), \\
$B_{c}=S_{c}(E_{1}= 0.30872\, {\rm MeV})$  & $260.2$ & $1.0$  & $13.6$ & $99.7$ &  $9.0$     \\   [2mm]
%\vspace{0.4cm}
\hline \hline \\
\end{tabular}
\end{center}
\end{table}
%\vspace{0.4cm}

\begin{table}
\begin{center}
\caption{${\tilde \chi}^{2}$ and $S(0)$ astrophysical factors for the ${}^{15}{\rm N}(p,\,\gamma){}^{16}{\rm O}$ capture process obtained from the unconstrained fit A(70) and constrained fit B(70).
\label{table_chiS70}} 
\begin{tabular}{ccc}
%[t]
\hline \hline \\
Fits & \,\, ${\tilde \chi}^{2}$\, \,&\,\, $S(0)$ [keVb] \\
\hline\\[1mm]\\ 
Fit A(70), $\,B_{c}=S_{c}(E_{1}=0.30872 {\rm MeV})$\,\,&\,\, $1.50$ &\,\,  $37.7 $ \\
\hline\\
Fit B(70), $\,B_{c}=S_{c}(E_{1}= 0.30872 {\rm MeV})$ \,\,    &\,\, $ 1.51$ &  $37.1 $\\
[2mm]
%\vspace{0.4cm}
\hline \hline \\
\end{tabular}
\end{center}
\end{table}

\section{Summary}
Determination of the $S(0)$ factor for the ${}^{15}{\rm N}(p,\,\gamma){}^{16}{\rm O}$ radiative capture is one of the goal of our fits. Although new measurements of this reaction \cite{LeBlanc} is a real success and a very important contribution to study of this reaction, we believe that it would be difficult to give a more accurate $S(0)$ value than the range  $33.1 \leq S(0) \leq 40.1$ keVb determined from our fits without further measurements down to lower energies than those achieved in \cite{LeBlanc}. To get more accurate uncertainties of the $S(0)$ factor a better estimate of energy uncertainties would be also useful. From our fits we determine the interval of the astrophysical factors at the effective energy $E=23.44$ keV 
$\,\,36.0\,{\rm keVb} \leq S(E=23.44 \,{\rm keV}) \leq  44.46\,{\rm keVb}$. Assuming the astrophysical factor 
$84.1 \pm 5.9$ MeVb  for the competing reaction ${}^{15}{\rm N}(p,\,\alpha){}^{12}{\rm C}$ \cite{lacognata09}
we find that for every $2084^{+413}_{-326}$ cycles of the main CN cycle one CN catalyst is lost due to the 
${}^{15}{\rm N}(p,\,\gamma){}^{16}{\rm O}$ reaction. 

But what is even more important is the question whether the minimum of ${\tilde \chi}^{2}$ is always an acceptable fit. 
Definitely, the answer is yes if our knowledge about the reaction model is complete. But it is assumed that the best fit is achieved under constrained variations of the fitting parameters within the accepted boundaries
obtained from the available physical information. This question elevates when the input physics is not complete.
It is another goal of our analysis to demonstrate that, due to our incomplete knowledge of the reaction model, it is not canonical that a fit, which provides minimum of ${\tilde \chi}^{2}$, is the best from the point of view of physics. We have demonstrated here that it is possible to achieve the same or even better fit and similar final $S(0)$ factors as in \cite{LeBlanc} by adopting the ANC measured from the transfer reaction rather then using an unconstrained variation of the ANC. But, even if the ANC is fixed within the experimental boundaries, the question remains about the interpretation of other fitting parameters. We have demonstrated problems with the interpretation of the parameters of the fits A and fit in \cite{LeBlanc}. Trying to improve interpretation we fixed some parameters, which are available from the analysis of the ${}^{15}{\rm N}(p,\,\alpha){}^{12}{\rm C}$ reaction, and  we are able to achieve even better fit than in  \cite{LeBlanc} and better agreement of the fitting parameters with the previous measurements of the ${}^{15}{\rm N}(p,\,\gamma){}^{16}{\rm O}$,  $\,\,{}^{15}{\rm N}(p,\,\alpha){}^{12}{\rm C}$ and  ${}^{15}{\rm N}(p,\,p){}^{15}{\rm N}$ processes \cite{TUNL}. However, still even our constrained fits are not fully satisfactory because we got too high value of the radiative width of the second resonance if we assume that the radiative width of the first resonance is $\sim 10$ eV. This issue remains to be resolved.

\section{Acknowledgments}
   
%We thanks Dr. D. Bemmerer for presenting the data from \cite{LeBlanc}. 
The work was supported by the US Department of Energy under Grants No. DE-FG02-93ER40773 and NSF under Grant No. PHY-0852653.
M.L. acknowledges the support by the Italian Ministry of University and Research under Grant No. RBFR082838 (FIRB2008).

\bibliography{15Npgbriefcommun01072010}

\begin{thebibliography}{24}
\expandafter\ifx\csname natexlab\endcsname\relax\def\natexlab#1{#1}\fi
\expandafter\ifx\csname bibnamefont\endcsname\relax
  \def\bibnamefont#1{#1}\fi
\expandafter\ifx\csname bibfnamefont\endcsname\relax
  \def\bibfnamefont#1{#1}\fi
\expandafter\ifx\csname citenamefont\endcsname\relax
  \def\citenamefont#1{#1}\fi
\expandafter\ifx\csname url\endcsname\relax
  \def\url#1{\texttt{#1}}\fi
\expandafter\ifx\csname urlprefix\endcsname\relax\def\urlprefix{URL }\fi
\providecommand{\bibinfo}[2]{#2}
\providecommand{\eprint}[2][]{\url{#2}}

\bibitem[{\citenamefont{Mukhamedzhanov et~al.}(2008)}]{muk08}
\bibinfo{author}{\bibfnamefont{A.~M.} \bibnamefont{Mukhamedzhanov}}
  \bibnamefont{et~al.}, \bibinfo{journal}{Phys.\ Rev.\ C}
  \textbf{\bibinfo{volume}{78}}, \bibinfo{pages}{015804}
  (\bibinfo{year}{2008}).

\bibitem[{\citenamefont{Rolfs and Rodney}(1974)}]{rolfs74}
\bibinfo{author}{\bibfnamefont{C.}~\bibnamefont{Rolfs}} \bibnamefont{and}
  \bibinfo{author}{\bibfnamefont{W.~S.} \bibnamefont{Rodney}},
  \bibinfo{journal}{Nucl.\ Phys.\ A} \textbf{\bibinfo{volume}{235}},
  \bibinfo{pages}{450} (\bibinfo{year}{1974}).

\bibitem[{\citenamefont{Hebbard}(1960)}]{hebb60}
\bibinfo{author}{\bibfnamefont{D.~F.} \bibnamefont{Hebbard}},
  \bibinfo{journal}{Nucl.\ Phys.} \textbf{\bibinfo{volume}{15}},
  \bibinfo{pages}{289} (\bibinfo{year}{1960}).

\bibitem[{\citenamefont{Barker}(2008{\natexlab{a}})}]{barker08}
\bibinfo{author}{\bibfnamefont{F.~C.} \bibnamefont{Barker}},
  \bibinfo{journal}{Phys.\ Rev.\ C} \textbf{\bibinfo{volume}{78}},
  \bibinfo{pages}{044611} (\bibinfo{year}{2008}{\natexlab{a}}).

\bibitem[{\citenamefont{Bemmerer et~al.}(2009)}]{bemmerer}
\bibinfo{author}{\bibfnamefont{D.}~\bibnamefont{Bemmerer}}
  \bibnamefont{et~al.}, \bibinfo{journal}{Journal of Physics G: Nuclear
  Physics} \textbf{\bibinfo{volume}{36}}, \bibinfo{pages}{045202}
  (\bibinfo{year}{2009}).

\bibitem[{\citenamefont{LeBlanc et~al.}(2010)}]{LeBlanc}
\bibinfo{author}{\bibfnamefont{P.}~\bibnamefont{LeBlanc}} \bibnamefont{et~al.},
  \bibinfo{journal}{Phys.\ Rev.\ C} \textbf{\bibinfo{volume}{82}},
  \bibinfo{pages}{055804} (\bibinfo{year}{2010}).

\bibitem[{\citenamefont{Blokhintsev et~al.}(1977)\citenamefont{Blokhintsev,
  Borbely, and Dolinskii}}]{blokhintsev}
\bibinfo{author}{\bibfnamefont{L.~D.} \bibnamefont{Blokhintsev}},
  \bibinfo{author}{\bibfnamefont{I.}~\bibnamefont{Borbely}}, \bibnamefont{and}
  \bibinfo{author}{\bibfnamefont{E.~I.} \bibnamefont{Dolinskii}},
  \bibinfo{journal}{Fiz. Elem. Chastits At. Yadra}
  \textbf{\bibinfo{volume}{8}}, \bibinfo{pages}{1189} (\bibinfo{year}{1977}).

\bibitem[{\citenamefont{Mukhamedzhanov et~al.}(2003)}]{muk2003}
\bibinfo{author}{\bibfnamefont{A.~M.} \bibnamefont{Mukhamedzhanov}}
  \bibnamefont{et~al.}, \bibinfo{journal}{Phys.\ Rev.\ C}
  \textbf{\bibinfo{volume}{67}}, \bibinfo{pages}{065804}
  (\bibinfo{year}{2003}).

\bibitem[{\citenamefont{Adelberger et~al.}(2010)}]{solarfusion}
\bibinfo{author}{\bibfnamefont{E.~G.} \bibnamefont{Adelberger}}
  \bibnamefont{et~al.} (\bibinfo{year}{2010}), \eprint{arXiv:1004.2318}.

\bibitem[{\citenamefont{La~Cognata et~al.}(2009)}]{lacognata09}
\bibinfo{author}{\bibfnamefont{M.}~\bibnamefont{La~Cognata}}
  \bibnamefont{et~al.}, \bibinfo{journal}{Phys.\ Rev.\ C}
  \textbf{\bibinfo{volume}{80}}, \bibinfo{pages}{012801(R)}
  (\bibinfo{year}{2009}).

\bibitem[{\citenamefont{Redder et~al.}(1982)}]{rolfs82}
\bibinfo{author}{\bibfnamefont{A.}~\bibnamefont{Redder}} \bibnamefont{et~al.},
  \bibinfo{journal}{Z. Phys. A} \textbf{\bibinfo{volume}{305}},
  \bibinfo{pages}{325} (\bibinfo{year}{1982}).

\bibitem[{\citenamefont{Azuma et~al.}(2010)}]{azuma}
\bibinfo{author}{\bibfnamefont{R.}~\bibnamefont{Azuma}} \bibnamefont{et~al.},
  \bibinfo{journal}{Phys.\ Rev.\ C} \textbf{\bibinfo{volume}{81}},
  \bibinfo{pages}{015804} (\bibinfo{year}{2010}).

\bibitem[{\citenamefont{Dolinsky and Mukhamedzhanov}(1977)}]{muk77}
\bibinfo{author}{\bibfnamefont{E.~I.} \bibnamefont{Dolinsky}} \bibnamefont{and}
  \bibinfo{author}{\bibfnamefont{A.~M.} \bibnamefont{Mukhamedzhanov}},
  \bibinfo{journal}{Izv. AN. SSSR, Ser. Fiz.} \textbf{\bibinfo{volume}{41}},
  \bibinfo{pages}{2055} (\bibinfo{year}{1977}).

\bibitem[{\citenamefont{Tang et~al.}(2003)}]{tang}
\bibinfo{author}{\bibfnamefont{X.}~\bibnamefont{Tang}} \bibnamefont{et~al.},
  \bibinfo{journal}{Phys.\ Rev.\ C} \textbf{\bibinfo{volume}{67}},
  \bibinfo{pages}{015804} (\bibinfo{year}{2003}).

\bibitem[{\citenamefont{Leuschner et~al.}(1994)}]{Leuschner}
\bibinfo{author}{\bibfnamefont{M.~B.} \bibnamefont{Leuschner}}
  \bibnamefont{et~al.}, \bibinfo{journal}{Phys.\ Rev.\ C}
  \textbf{\bibinfo{volume}{49}}, \bibinfo{pages}{955} (\bibinfo{year}{1994}).

\bibitem[{\citenamefont{Kramer et~al.}(2001)\citenamefont{Kramer, Blokb, and
  Lapiks}}]{kramer}
\bibinfo{author}{\bibfnamefont{G.}~\bibnamefont{Kramer}},
  \bibinfo{author}{\bibfnamefont{H.}~\bibnamefont{Blokb}}, \bibnamefont{and}
  \bibinfo{author}{\bibfnamefont{L.}~\bibnamefont{Lapiks}},
  \bibinfo{journal}{Nucl.\ Phys.\ A} \textbf{\bibinfo{volume}{679}},
  \bibinfo{pages}{267} (\bibinfo{year}{2001}).

\bibitem[{\citenamefont{Hiebert et~al.}(1967)\citenamefont{Hiebert, Newman, and
  Bassel}}]{Hiebert}
\bibinfo{author}{\bibfnamefont{J.}~\bibnamefont{Hiebert}},
  \bibinfo{author}{\bibfnamefont{E.}~\bibnamefont{Newman}}, \bibnamefont{and}
  \bibinfo{author}{\bibfnamefont{R.}~\bibnamefont{Bassel}},
  \bibinfo{journal}{Phys.\ Rev.} \textbf{\bibinfo{volume}{154}},
  \bibinfo{pages}{898} (\bibinfo{year}{1967}).

\bibitem[{\citenamefont{Geurts et~al.}(1996)}]{Geurts}
\bibinfo{author}{\bibfnamefont{W.~J.~W.} \bibnamefont{Geurts}}
  \bibnamefont{et~al.}, \bibinfo{journal}{Phys.\ Rev.\ C}
  \textbf{\bibinfo{volume}{53}}, \bibinfo{pages}{2207} (\bibinfo{year}{1996}).

\bibitem[{\citenamefont{Jensen et~al.}(1996)}]{Jensen}
\bibinfo{author}{\bibfnamefont{H.}~\bibnamefont{Jensen}} \bibnamefont{et~al.},
  \bibinfo{journal}{Phys.\ Rev.\ C} \textbf{\bibinfo{volume}{82}},
  \bibinfo{pages}{014310} (\bibinfo{year}{1996}).

\bibitem[{\citenamefont{Lane and Thomas}(1958)}]{lanethomas58}
\bibinfo{author}{\bibfnamefont{A.~M.} \bibnamefont{Lane}} \bibnamefont{and}
  \bibinfo{author}{\bibfnamefont{R.~G.} \bibnamefont{Thomas}},
  \bibinfo{journal}{Rev. Mod. Phys.} \textbf{\bibinfo{volume}{30}},
  \bibinfo{pages}{257} (\bibinfo{year}{1958}).

\bibitem[{\citenamefont{Barker and Kajino}(1991)}]{barkerkajino}
\bibinfo{author}{\bibfnamefont{F.~C.} \bibnamefont{Barker}} \bibnamefont{and}
  \bibinfo{author}{\bibfnamefont{T.}~\bibnamefont{Kajino}},
  \bibinfo{journal}{Aust. J. Phys.} \textbf{\bibinfo{volume}{44}},
  \bibinfo{pages}{369} (\bibinfo{year}{1991}).

\bibitem[{\citenamefont{Barker}(1972)}]{barker72}
\bibinfo{author}{\bibfnamefont{F.~C.} \bibnamefont{Barker}},
  \bibinfo{journal}{Aust. J. Phys.} \textbf{\bibinfo{volume}{25}},
  \bibinfo{pages}{341} (\bibinfo{year}{1972}).

\bibitem[{\citenamefont{Barker}(2008{\natexlab{b}})}]{barker2008}
\bibinfo{author}{\bibfnamefont{F.~C.} \bibnamefont{Barker}},
  \bibinfo{journal}{Phys.\ Rev.\ C} \textbf{\bibinfo{volume}{78}},
  \bibinfo{pages}{044612} (\bibinfo{year}{2008}{\natexlab{b}}).

\bibitem[{\citenamefont{Tilley et~al.}(1993)\citenamefont{Tilley, Weller, and
  Cheves}}]{TUNL}
\bibinfo{author}{\bibfnamefont{D.~R.} \bibnamefont{Tilley}},
  \bibinfo{author}{\bibfnamefont{H.~R.} \bibnamefont{Weller}},
  \bibnamefont{and} \bibinfo{author}{\bibfnamefont{C.~M.}
  \bibnamefont{Cheves}}, \bibinfo{journal}{Nucl.\ Phys.\ A}
  \textbf{\bibinfo{volume}{565}}, \bibinfo{pages}{1} (\bibinfo{year}{1993}).

\end{thebibliography}
\end{document}